\documentclass{PoS}

\usepackage{epsfig}
\usepackage{cite}

\def\({\left(}
\def\){\right)}
\def\[{\left[}
\def\]{\right]}

\relax

\def    \hepph  #1 {{\tt hep-ph/#1}}
\def    \hepex  #1 {{\tt hep-ex/#1}}

\def\Black{}

\newsavebox\tmpfig

\newcommand{\lp}{\left(}
\newcommand{\rp}{\right)}
\newcommand{\be}{\begin{equation}}
\newcommand{\ee}{\end{equation}}

\title{The impact of
heavy quark mass effects in the NNPDF
global analysis}

\ShortTitle{Heavy quark mass effects in NNPDF}

\author{\speaker{J. Rojo}, S. Forte\\
         Dipartimento di Fisica, Universit\`a di Milano and
INFN, Sezione di Milano,\\ Via Celoria 16, I-20133 Milano, Italy\\
        E-mail: \email{juan.rojo@mi.infn.it}, \email{stefano.forte@mi.infn.it}
}

\author{R. D. Ball. L. Del Debbio, M. Ubiali\\
         School of Physics and Astronomy, University of Edinburgh,\\
JCMB, KB, Mayfield Rd, Edinburgh EH9 3JZ, Scotland\\
        E-mail: \email{rdb@ph.ed.ac.uk}, 
\email{luigi.del.debbio@ed.ac.uk}, \email{maria.ubiali@gmail.com}
}

\author{V. Bertone, A. Guffanti\\
         Physikalisches Institut, Albert-Ludwigs-Universit\"at Freiburg
\\ Hermann-Herder-Stra\ss e 3, D-79104 Freiburg i. B., Germany \\
        E-mail: \email{Valerio.Bertone@physik.uni-freiburg.de}, 
\email{alberto.guffanti@physik.uni-freiburg.de}
}

\author{F. Cerutti, J. I. Latorre\\
        Departament d'Estructura i Constituents de la Mat\`eria, 
Universitat de Barcelona,\\ Diagonal 647, E-08028 Barcelona, Spain\\
        E-mail: \email{francesco.cerutti@gmail.com}, \email{latorre@ecm.ub.es}
}

\abstract{We discuss the implementation of the 
FONLL general-mass scheme for
heavy quarks in deep-inelastic scattering in the FastKernel framework,
used in the NNPDF series of global PDF analysis. 
We present the general features of FONLL and benchmark
the accuracy of its implementation in FastKernel comparing
with the Les Houches heavy quark benchmark tables.
We then show preliminary
results of the NNPDF2.1 analysis, in which 
heavy quark mass effects are included following
the FONLL-A GM scheme.}

\FullConference{DIS 2010}

\begin{document}

\section{FONLL in Mellin space: FastKernel}

The NNPDF series of 
PDF analysis~\cite{DelDebbio:2007ee,Ball:2008by,Ball:2009mk,Ball:2009qv,Ball:2010de}
has used up to now the ZM-VFN scheme for the treatment of heavy flavours.
In this contribution we review progress on the implementation
of heavy quark mass effects in the NNPDF analysis and their
implications on PDFs and LHC observables. We present preliminary
results of the NNPDF2.1 set~\cite{nnpdf21}, which
updates the global NNPDF2.0 fit with heavy quark mass effects
 as well as with the addition
of the H1 and ZEUS charm structure function data $F_2^c$.

Recently, the FONLL general-mass scheme, originally formulated
for heavy quark hadroproduction, was generalized
to deep-inelastic scattering~\cite{Forte:2010ta}. This scheme has
several advantages over other existing GM schemes, such as
S-ACOT~\cite{Aivazis:1993pi}, 
TR/MSTW08~\cite{Thorne:2006qt} and BMSN~\cite{Buza:1996wv}: it can be
applied also in general hadronic processes, 
it can be formulated at any perturbative order, 
 it allows the combination of different
perturbative orders in the massless and massive
computations, and it can be combined with various prescriptions
for the treatment of subleading mass-suppressed terms near threshold,
such as $\chi$--scaling or damping factors. 
In Ref.~\cite{Forte:2010ta}
explicit FONLL implementations are discussed 
at $\mathcal{O}\lp \alpha_s\rp$ (FONLL-A) and
 at $\mathcal{O}\lp \alpha_s^2\rp$, either within an NLO 
(FONLL-B) or within a NNLO calculation
(FONLL-C).  

\begin{figure}[ht]
\begin{center}
\epsfig{file=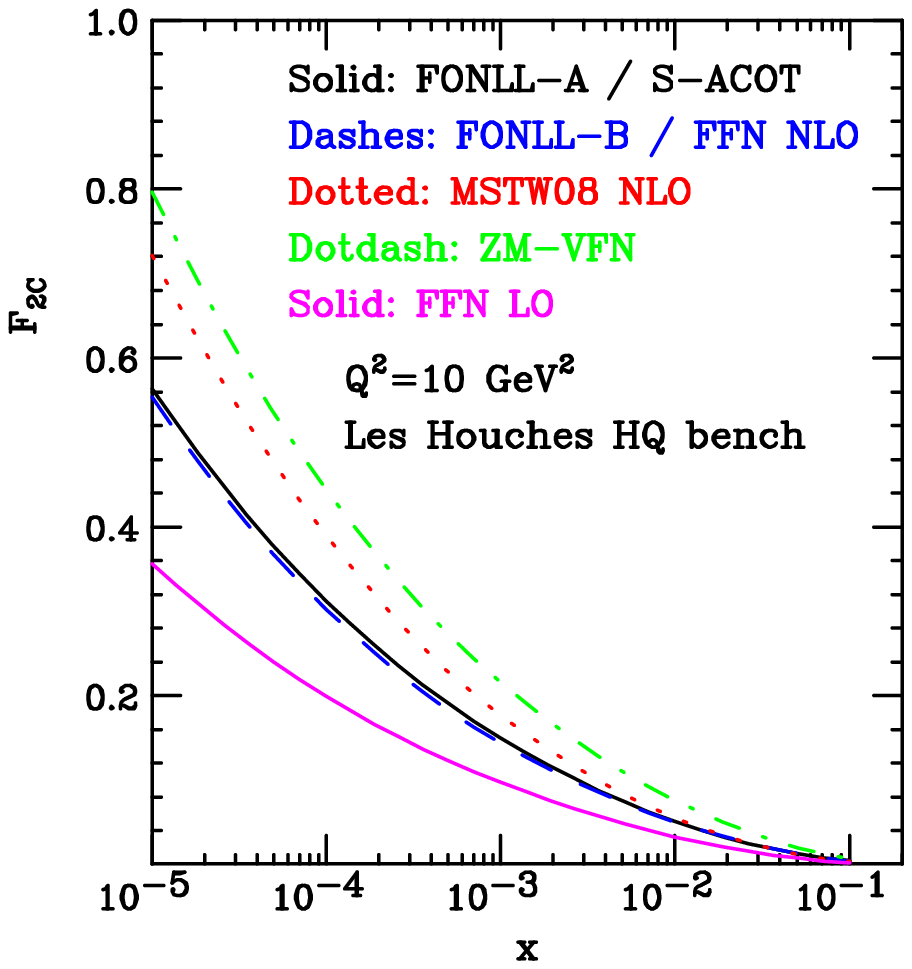,width=0.45\textwidth,angle=90}
\epsfig{file=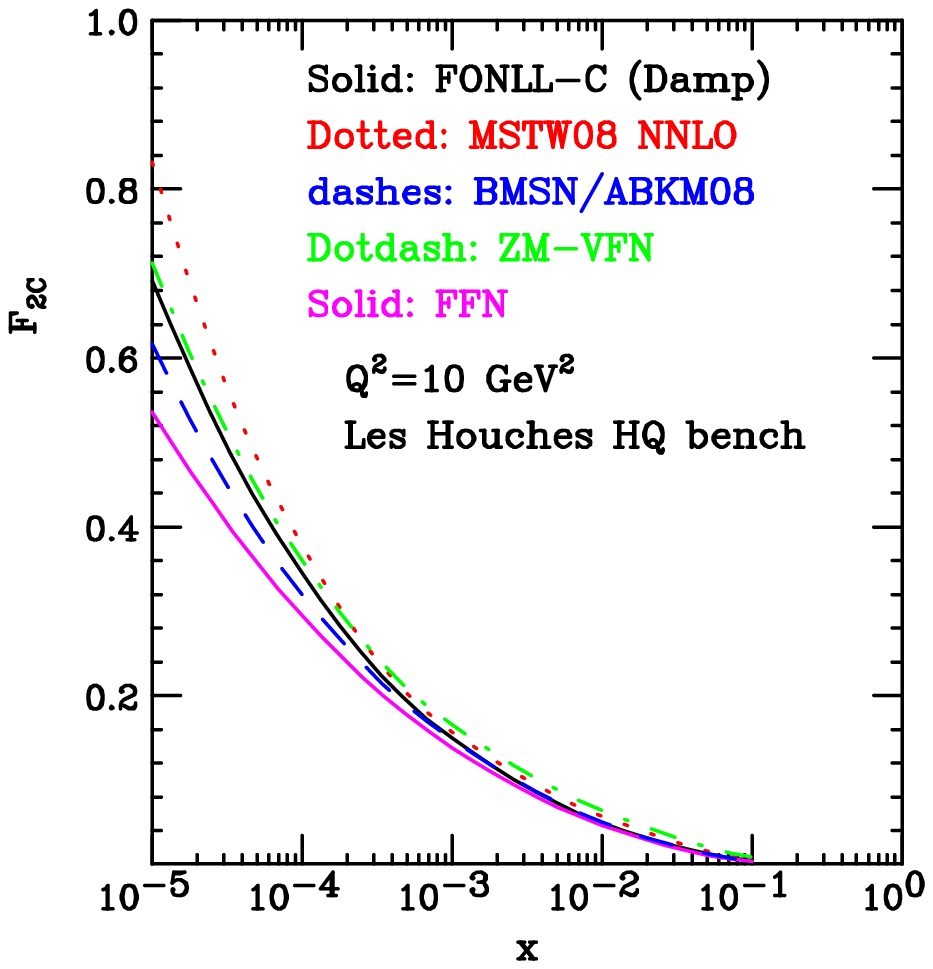,width=0.45\textwidth,angle=90}
\caption{\small 
\label{LH_HQ_fig:F2c-comp-talk} The $F_{2}^c$ structure function for $Q^2=10$
GeV$^2$ computed at $\mathcal{O}\lp \alpha_s\rp$
accuracy (left plot) and
$\mathcal{O}\lp \alpha_s^2\rp$ accuracy (right plot) 
in various GM schemes: FONLL, S-ACOT, MSTW08 and BMSN.
The purely massless and massive results are also shown
for comparison.}
\end{center}
\end{figure}

The different
versions of FONLL were compared to other commonly used GM schemes
in the Les Houches heavy quark benchmark study~\cite{Binoth:2010ra}. 
Fig.~\ref{LH_HQ_fig:F2c-comp-talk} summarizes the result of this
comparison. At $\mathcal{O}\lp \alpha_s\rp$, FONLL-A and S-ACOT
are formally identical, while FONLL-B cannot be distinguished from
the $\mathcal{O}\lp \alpha_s^2\rp$ massive result at this value
of $Q^2$.
In  Fig.~\ref{LH_HQ_fig:F2c-comp-talk}, the S-ACOT and FONLL-A  
curves have been obtained using a threshold damping 
factor prescription~\cite{Forte:2010ta}, the results with 
 $\chi$--scaling are very close numerically. The MSTW08 result
instead has been obtained using the $\chi_2$ version of the $\chi-$scaling
prescription~\cite{Binoth:2010ra}, which is rather different near threshold.
The purely massless and massive results are also shown
for illustration. It can be seen that all general-mass schemes
interpolate between the massless and massive calculations, and that
there are significant differences between them, whose impact
in PDF determinations is yet still to be systematically explored.

\begin{figure}[t!]
\begin{center}
\includegraphics[width=0.99\textwidth]{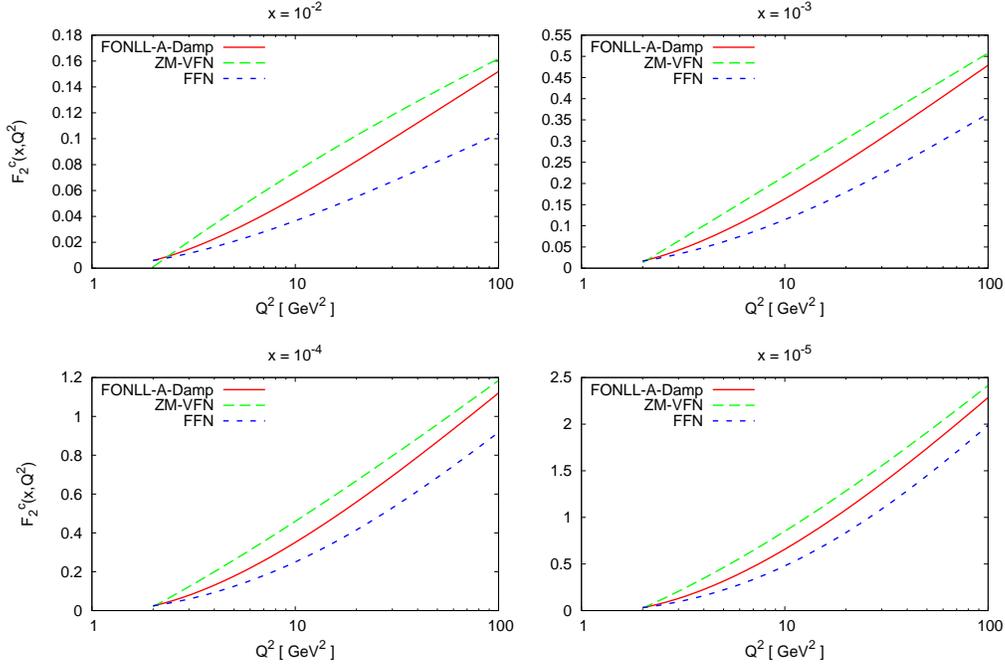}
\end{center}
\caption{\small The charm structure functions $F_2^c(x,Q^2)$ 
as a function of $Q^2$ for different values of $x$ from
$x=10^{-5}$ to $x=10^{-2}$ in various 
heavy quark schemes, computed using the FastKernel method:
FONLL-A-Damp, ZM-VFN and the FFN scheme. The
PDFs and settings are identical to those of the Les Houches
Heavy quark benchmark comparison~\cite{Binoth:2010ra}.}
\label{fig:f2c_vs_q2}
\end{figure}

The FONLL GM scheme has been now implemented in the
NNPDF global analysis framework. 
As discussed in Refs.~\cite{Ball:2008by,Ball:2010de}, for the PDF
evolution and the computation of physical observables the NNPDF analysis is 
based on the FastKernel 
framework, which allows a fast and accurate PDF evolution
and evaluation of physical observables, including hadronic
processes, without any K--factor approximation at all. 
The implementation of heavy quark
effects in FastKernel has required the computation of the Mellin
transforms of the $\mathcal{O}\lp \alpha_s\rp$ massive DIS
coefficient functions, both in the Neutral Current
and in the Charged Current sector. These are not currently 
available in closed
form; complete
analytic expressions will be presented in Ref.~\cite{nnpdf21},
here we limit ourselves of presenting the results for the
accuracy using the Les Houches heavy quark benchmark settings.
We summarize the results of the comparison
for $F_2^c$ in Table~\ref{tab:tablebench}. In all cases the accuracy is well
below the percent level, accurate enough for precision phenomenology.
In Fig.~\ref{fig:f2c_vs_q2} we compare the $F_2^c$ structure
function for various heavy quark schemes: ZM, FONLL-A-Damp
and the FFN scheme as a function of $Q^2$ for different values of
$x$. It can easily be seen how FONLL-A-Damp smoothly interpolates
between the FFN scheme near threshold and the massless scheme
at large $Q^2$.

\begin{table}[ht]
\begin{center}
\scriptsize
 \begin{tabular}{|c|c|c|c|}
\hline
& \multicolumn{3}{|c|}{FONLL-A-Damp}  \\
 \hline
 $x$ & FONLLdis & FastKernel  
& Accuracy \\
\hline
 \hline
\multicolumn{4}{|c|}{$Q^2=4$ GeV$^2$} \\
 \hline
$10^{-5}$ &      0.1507  & 0.1501   & $0.4\%$ \\
 $10^{-4}$  &    0.0936  & 0.0931   & $0.5\%$  \\
 $10^{-3}$  &    0.0506   & 0.0504  & $0.4\%$ \\
  $10^{-2}$  &   0.0174   & 0.0176 & $0.9\%$  \\
\hline
\hline
\multicolumn{4}{|c|}{$Q^2=10$ GeV$^2$} \\
 \hline
$10^{-5}$ &    0.563 & 0.561 & $0.4\%$ \\
 $10^{-4}$  &  0.312 & 0.311 & $0.3\%$ \\
 $10^{-3}$  &   0.1499 & 0.1495 & $0.3\%$\\
  $10^{-2}$  &  0.05056 & 0.05052 & $0.1\%$ \\
\hline
\hline
\multicolumn{4}{|c|}{$Q^2=100$ GeV$^2$} \\
 \hline
$10^{-5}$ &    2.28636 & 2.28577 & $0.02\%$  \\
 $10^{-4}$  &  1.12186 & 1.12082 & $0.1\%$  \\
 $10^{-3}$  &  0.48008 & 0.47919 & $0.2\%$  \\
  $10^{-2}$  & 0.15207 & 0.15200 & $0.04\%$ \\
 \hline
 \end{tabular}
\end{center}
\caption{\small Results of the benchmark comparison for the
$F_{2c}(x,Q^2)$ structure function in the
 FONLL-A-Damp scheme for the FONLLdis code~\cite{Forte:2010ta}
and for the FastKernel framework. Results are provided at the
benchmark kinematical points in $x,Q^2$~\cite{Binoth:2010ra}.
\label{tab:tablebench}}
\end{table}

\section{Heavy quark mass effects in the NNPDF global analysis}

\begin{figure}[t!]
\begin{center}
\includegraphics[width=0.47\textwidth]{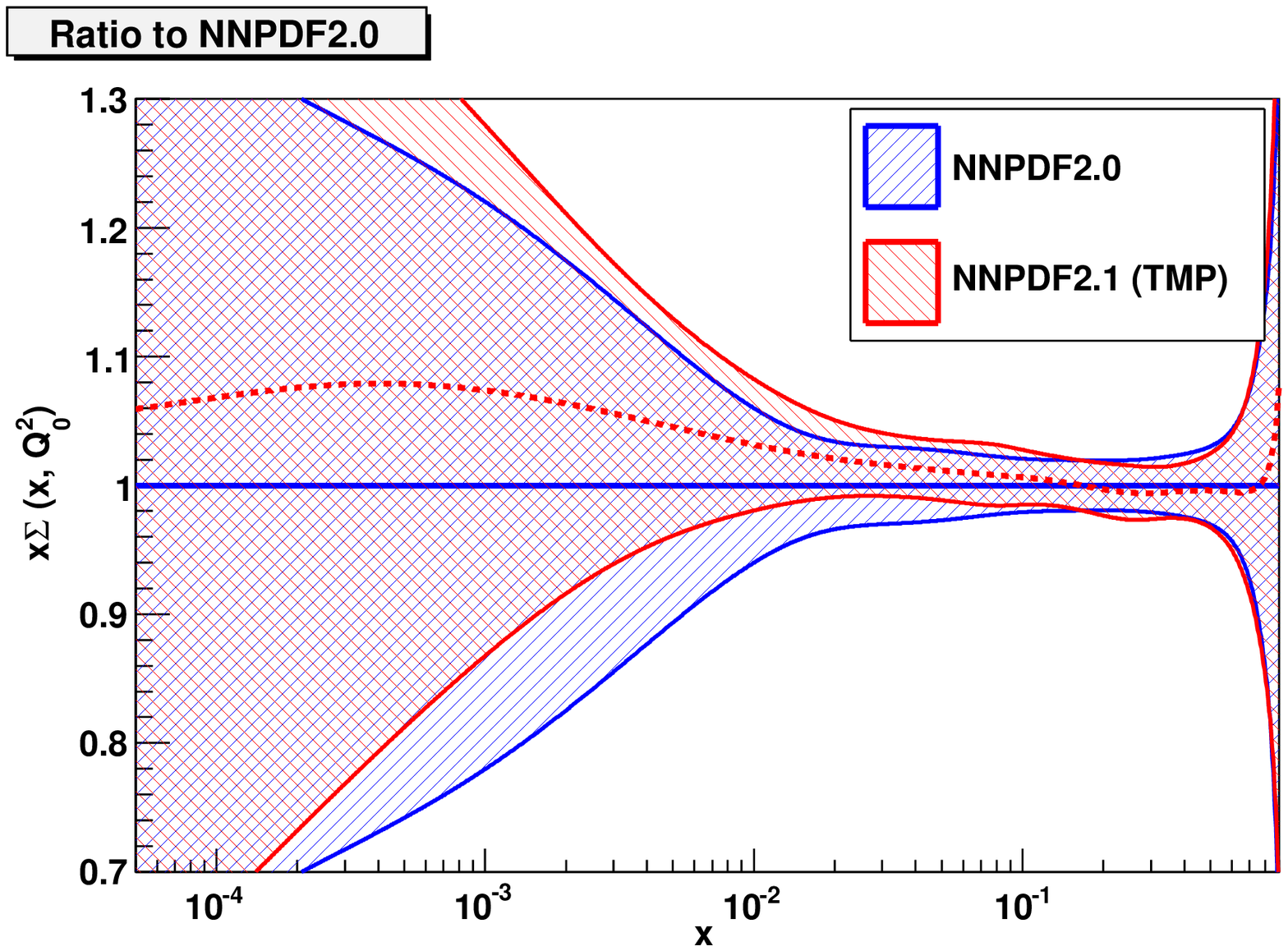}
\includegraphics[width=0.47\textwidth]{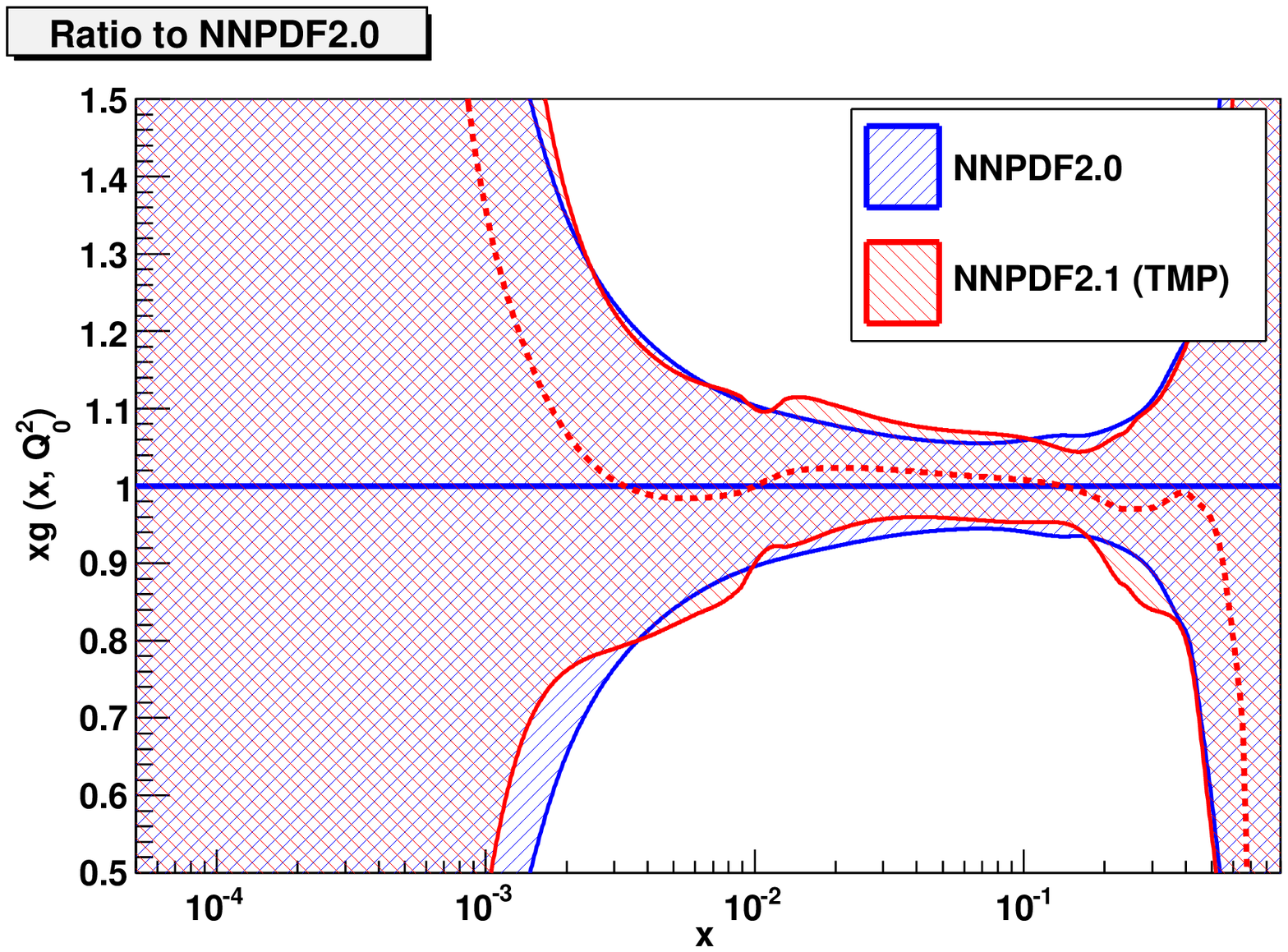}
\includegraphics[width=0.47\textwidth]{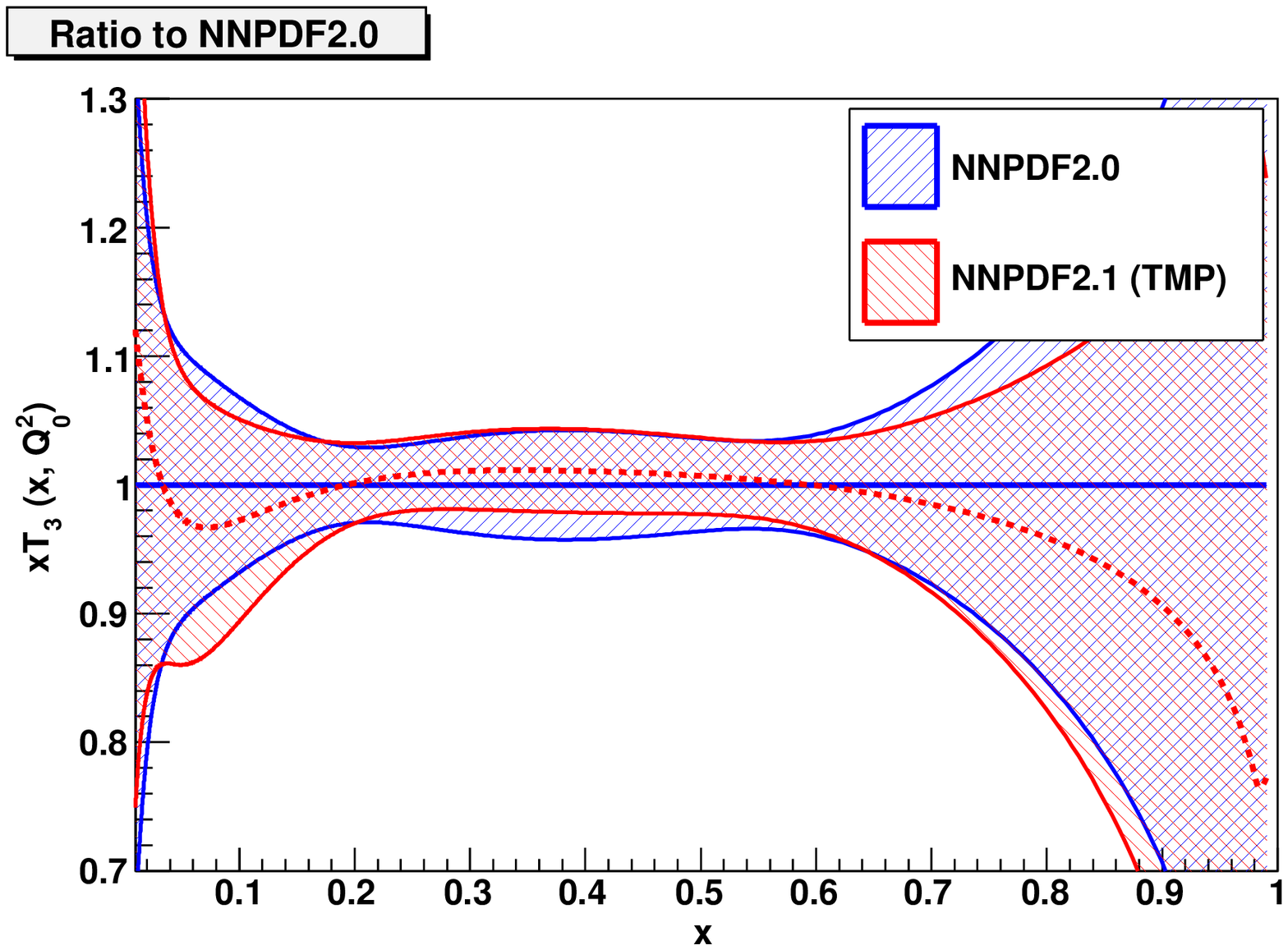}
\includegraphics[width=0.47\textwidth]{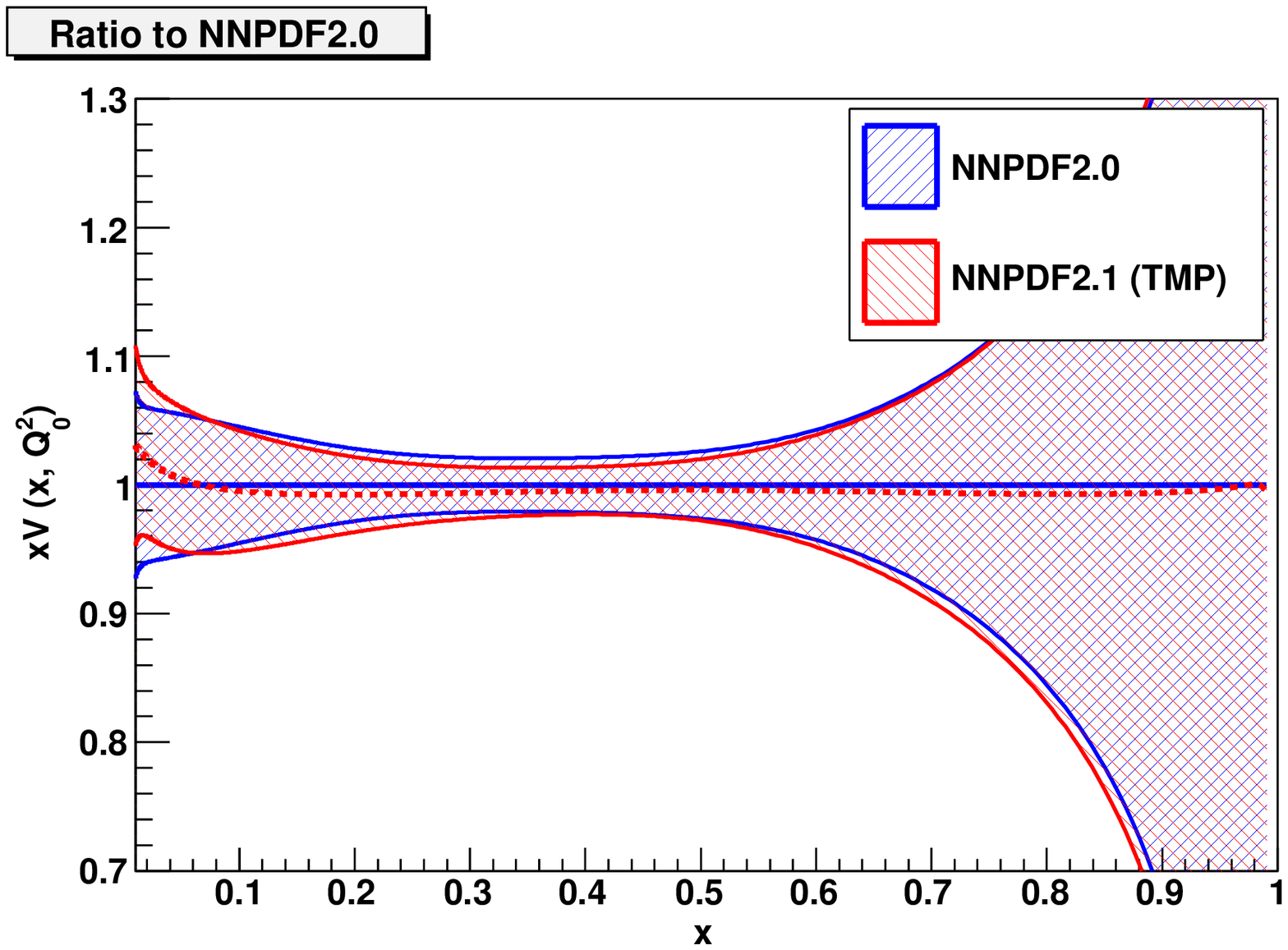}
\end{center}
\caption{\small Upper plots: The ratio of the preliminary NNPDF2.1 Singlet
(left plot) and gluon PDFs (right plot) to the respective NNPDF2.0
ones, at the initial evolution scale
$Q_0^2=2$ GeV$^2$. Error bands are normalized to the NNPDF2.0 central value.
Lower plots: same as above for the triplet $T_3$ and the total
valence $V$.}
\label{fig:nnpdf21comp}
\end{figure}

Here we present preliminary results based on a fit to the
 NNPDF2.0 dataset~\cite{Ball:2010de}, supplemented
by charm structure 
function data $F_2^c(x,Q^2)$
from the H1 and ZEUS experiments at HERA: NNPDF2.1. 
In Fig.~\ref{fig:nnpdf21comp} we perform
a preliminary comparison of the singlet and gluon PDFs
at the initial evolution scale $Q_0^2=2$ GeV$^2$ in NNPDF2.1
and NNPDF2.0, normalized to the NNPDF2.0 central values. 
As expected, including heavy quark mass effects leads
to a increase in the singlet at medium and small-$x$,
as well as to a marked increase in the small-$x$ gluon. Note however that
differences are always within the PDF uncertainty bands. 
This implies that heavy quark mass effects should modify the
NNPDF2.0 predictions for LHC observables by 1--sigma or so 
at most. The implications of these results
for LHC observables 
will be discussed in detail in Ref.~\cite{nnpdf21}.
 This is to be compared
with the CTEQ analysis~\cite{Tung:2006tb}, where for example
a $\sim 2.5$--sigma variation on $\sigma_W$ was obtained
in the GM fit based on S-ACOT-$\chi$ as compared to the
ZM case.

We also show in 
Fig.~\ref{fig:nnpdf21comp} the effects of heavy quark mass
effects in the triplet and total valence PDFs: as expected their
impact is completely negligible. Since processes which depend
on valence PDFs, as well as those on the the medium and large-$x$
gluon like Higgs production~\cite{Demartin:2010er}, are
 unaffected by heavy quark mass effects, the predictions for such
processes  obtained
with NNPDF2.0 will be very close to those
of NNPDF2.1.

\end{document}